\documentstyle[epsf] {elsart}

\begin {document}
\begin{frontmatter}

\title {The CAT Imaging Telescope
for Very-High-Energy Gamma-Ray Astronomy}

\author[paris]{A. Barrau}
\author[cesr]{R. Bazer-Bachi}
\author[x]{E. Beyer}
\author[perpi]{H. Cabot\thanksref{dead}}
\author[x]{M. Cerutti}
\author[x]{L.M. Chounet}
\author[perpi]{G. Debiais}
\author[x]{B. Degrange}
\author[paris]{H. Delchini}
\author[paris]{J.P. Denance}
\author[paris]{G. Descotes\thanksref{dead}}
\author[cesr]{J.P. Dezalay}
\author[paris]{A. Djannati-Ata\"{\i}}
\author[cenbg]{D. Dumora}
\author[cdf]{P. Espigat}
\author[perpi]{B. Fabre}
\author[x]{P. Fleury}
\author[x]{G. Fontaine}
\author[paris]{R. George}
\author[cdf]{C. Ghesqui\`ere}
\author[x]{J. Gilly}
\author[sap]{P. Goret}
\author[sap]{C. Gouiffes}
\author[cenbg]{J.C. Gouillaud}
\author[x]{C. Gr\'egory}
\author[sap,paris7]{I.A. Grenier}
\author[x]{L. Iacoucci}
\author[x]{L. Kalt}
\author[cdf]{S. Le Bohec}
\author[cesr]{I. Malet}
\author[perpi]{C. Meynadier}
\author[sap]{J.P. Mols}
\author[x]{P. Mora de Freitas}
\author[x]{R. Morano}
\author[x]{G. Morinaud}
\author[cdf,czech]{F. Munz}
\author[czopt]{M. Palatka}
\author[purd]{T.A. Palfrey}
\author[x]{E. Par\'e}
\author[paris]{Y. Pons}
\author[cdf]{M. Punch}
\author[cenbg]{J. Qu\'ebert}
\author[cenbg]{K. Ragan}
\author[sap,paris7]{C. Renault}
\author[paris]{M. Rivoal}
\author[czech]{L. Rob}
\author[czopt]{P. Schovanek}
\author[cenbg]{D. Smith}
\author[sap]{A. Tabary}
\author[paris]{J.P. Tavernet}
\author[paris]{F. Toussenel}
\author[x]{J. Vrana\thanksref{dead}}

\address [cenbg]{Centre d'Etudes Nucl\'eaire de
Bordeaux-Gradignan, France\thanksref{in2p3}}
\address [cesr]{Centre d'Etudes Spatiales des
Rayonnements, Toulouse, France\thanksref{insu}}
\address [cdf]{Laboratoire de Physique Corpuscul\-aire et Cosmologie,
Coll\`ege de France, Paris, France\thanksref{in2p3}}
\address [x]{Laboratoire de Physique Nucl\'eaire de Haute Energie,
Ecole Polytechnique, Palaiseau, France\thanksref{in2p3}}
\address [paris]{Laboratoire de Physique Nucl\'eaire de Haute Energie,
Universit\'es de Paris VI/VII, France\thanksref{in2p3}}
\address [sap]{Service d'Astrophysique, Centre d'Etudes
de Saclay, France\thanksref{cea}}
\address [perpi]{Groupe de Physique Fondamentale,
Universit\'e de Perpignan, France\thanksref{in2p3}}
\address [czech]{Nuclear Center, Charles University, Prague, Czech
Republic}
\address [paris7]{Universit\'e Paris VII, France}
\address [purd]{Department 
of Physics, Purdue University, Lafayette, IN
47907, U.S.A.}
\address [czopt]{Joint Laboratory 
of Optics Ac. Sci. and Palacky
University, Olomouc, Czech Republic}

\thanks [in2p3]{IN2P3/CNRS}
\thanks [insu]{INSU/CNRS}
\thanks [cea]{DAPNIA/CEA}
\thanks [dead]{Deceased}

\begin{abstract}
The CAT (Cherenkov Array at Th\'emis) imaging telescope, equipped with
a very-high-definition camera (546 fast phototubes with $0.12^{\circ}$
spacing surrounded by 54 larger tubes in two guard rings) started
operation in Autumn 1996 on the site of the former solar plant Th\'emis
(France). Using the atmospheric Cherenkov technique, it detects and
identifies very high energy $\gamma$-rays in the range $250\:{\mathrm
GeV}$ to a few tens of TeV. The instrument, which has detected three
sources (Crab nebula, Markarian 421 and Markarian 501), is described in
detail. 
\end{abstract}
\begin{keyword}
Gamma-Ray Astronomy, Atmospheric Cherenkov detector
\end{keyword}
\end{frontmatter}

\section { Introduction }
The recent development of Very-High-Energy Gamma-Ray Astronomy is
essentially due to Atmospheric Cherenkov Telescopes ({\small ACT}).
Since 1988, two arrays of Cherenkov detectors,  A{\small SGAT}
\cite{asgat} and T{\small HEMISTOCLE} \cite{themis} have been operated
on  the site of the former solar plant ``Th\'emis'', close to
Font-Romeu in the French Pyrenees. They consist of several stations
sampling the Cherenkov front on the ground and reconstructing the
direction of $\gamma$-ray showers through accurate timing measurements.
Besides this sampling technique, in which $\gamma$-ray showers from a
point-like source are discriminated from the large background due to
charged cosmic-rays by means of the angular resolution, an important
breakthrough was brought by the imaging technique, pioneered by the
Whipple group \cite{whitel}. In imaging Atmospheric Cherenkov
Telescopes, the image of the shower is formed on a multi-pixel camera
and the discrimination between $\gamma$-ray-induced and proton or
nucleus-induced showers is based both on image shape and
directionality. This technique proved to be the most powerful to reject
the background and led to the discovery of at least four firmly
established very-high-energy sources: the Crab nebula \cite{whicrab},
the pulsar {\small PSR} 1706-44 \cite{canpsr}, and the two relatively
nearby Active Galactic Nuclei, Markarian 421 \cite{whi421} and 
Markarian 501 \cite{whi501}. In 1993, it was therefore proposed to
complement the ``Th\'emis'' arrays by a high-performance imaging
telescope, the overall setup being called ``C{\small AT}'' for
``Cherenkov Array at Th\'emis''. The project was approved in 1994 and
the new imaging telescope was commissioned in September 1996.
Astrophysical observations started in October 1996. The main
characteristics of the new instrument are the following:
\begin{itemize}
\item The C{\small AT} imaging telescope achieves a relatively low
energy threshold ($\sim 250\:{\mathrm GeV}$) despite its moderate
reflector area ($\sim 18\:{\mathrm m}^2$). The threshold of an {\small
ACT} being essentially fixed by the  night-sky background, it is
important to protect phototubes from parasitic light by using Winston
cones  and to take full advantage of the rapidity of the Cherenkov
signal. This is achieved here by combining an almost isochronous
optics, fast phototubes with a good resolution at the single
photoelectron level, and fast trigger electronics located just behind
the photodetectors. By these means, the C{\small AT} imaging telescope
achieves a threshold similar to that of the Whipple {\small ACT}, whose
reflector area is $75\:{\mathrm m}^2$.
\item The C{\small AT} imaging telescope provides a very high image
definition (546 pixels with $0.12^{\circ}$ spacing in the central part
of the field of view), resulting in improved background rejection and
energy resolution. In a companion paper \cite{lebohec}, it is shown
that accurate analysis of the longitudinal light profile of the shower
image allows the direction of each $\gamma$-ray to be determined with a
resolution of the order of the pixel size, comparable to that of
H{\small EGRA} stereoscopic system \cite{hegra} at the same energies.
\item Particular care has been given to various monitoring facilities
concerning the mechanical structure and the optics, as well as the
phototubes. 
\item Finally, the ``Th\'emis'' arrays (Imaging Telescope,  A{\small
SGAT} and T{\small HEMIS\-TOCLE}) yield simultaneous measurements of
the energies of $\gamma$-ray showers by two complementary techniques:
imaging and Cherenkov-front sampling. Such data provide  a unique tool
for cross-calibration, of particular  importance for the determination
of source spectra. The T{\small HEMISTOCLE} array is now operated with
an energy threshold of $1.5\:{\mathrm TeV}$, whereas A{\small SGAT} (in
the process of being upgraded) will work in the low-energy range
($>300\:{\mathrm GeV}$). 
\end{itemize}

The general layout of the experiment is described in Section 2. Section
3 describes the optics and the mechanical structure and Section 4 the
focal plane detector.  The trigger and readout electronics are
explained in Section 5 and the on-line software in Section 6. 

\section { General setup }
\label{sec-setup}
\begin{figure}[htbp]
\epsfxsize=10cm
\centering
\leavevmode
\epsfbox{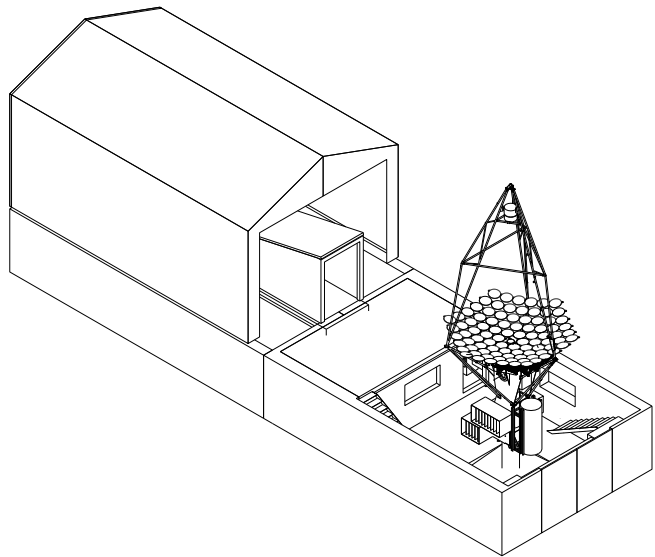}
\caption{A general view of the C{\small AT} imaging telescope with the
hangar opened. The hangar is oriented East-West, the telescope being on
the western side.}
\label{fig:han}
\end{figure}

The imaging telescope is located at latitude $42.50^{\circ}$N,
longitude $1.97^{\circ}$E and altitude $1650\:{\mathrm m}$. It uses the
alt-azimuthal mount of a former heliostat located approximately  at the
center of A{\small SGAT} and T{\small HEMISTOCLE} arrays.  Its
orientation is controlled by computer in steps of $0.008^{\circ}$ both
in elevation and azimuth. Its main mechanical and optical
characteristics are shown in Table~1. The instrument is sheltered from
bad weather by a building including a hangar mounted on rails which can
be rolled away during observing periods (Fig.~\ref{fig:han}). In the
garage position, the telescope axis is horizontal and the center of
curvature of the central mirror element is easily accessed from the
platform shown in Fig.~\ref{fig:han}. The setup used for mirror
orientation (see section 3.1) is located at this point and is sheltered
by the hut at the back of the platform when the hangar is closed. The
control room is located under the platform.
\begin{table}
\caption{Main mechanical and optical characteristics of the telescope}
\begin{tabbing}
Total reflector area:
................................................................\=
\kill
Total reflector area: \> $17.8\:{\mathrm m}^{2}$ \\
Effective reflector area: \> $16.0\:{\mathrm m}^{2}$ \\
Number of mirrors: \> $90$ \\
Mirror diameter: \> $0.50\:{\mathrm m}$ \\
Mirror curvature radius: \> $12\:{\mathrm m}$ \\
Focal length: \> $ 6\:{\mathrm m}$ \\
Total field of view (diameter): \> $4.8^{\circ}$ \\
Angular accuracy on mirror orientations: \> $< 0.01^{\circ}$ \\
Allowed range in elevation: \> $0^{\circ}$ to $90^{\circ}$ \\
Allowed range in azimuth: \> $0^{\circ}$ to $\pm 270^{\circ}$ \\
Possible rotation speed of the mount: \> $0.04^{\circ}{\mathrm s^{-1}}$ 
and $0.2^{\circ}{\mathrm s^{-1}}$ \\
Maximum wind speed for acceptable images: \> $32\:{\mathrm km/h}$ \\
Total weight of the moving part of the telescope (full load): \> $6\:
{\mathrm tons}$ 
\end{tabbing}
\end{table}

\section { Optical reflector and mechanical structure }

\subsection { The reflector }

The reflector is based on the Davies-Cotton design \cite{Davcot}. The
collecting area consists of $90$ spherical mirrors, $50\:{\mathrm cm}$
in diameter, all with approximately the same radius of curvature
$R=12\:{\mathrm m}$ (the measured values being in the range
$11.76\:{\mathrm m} < R < 12.0\:{\mathrm m}$), the centers of each
elementary reflecting dish being located on a sphere centered on the
reflector axis at a distance $R/2$ from the apex.
 
The mirror thickness ($\approx 10\:{\mathrm mm}$) was chosen as a
compromise between the weight (which must be accommodated by the mirror
support and the heliostat mount) and the optical accuracy.  Mirrors
were manufactured from borosilicate glass ({\small SIMAX}) using a
steel pressing form at a temperature of $1450^{\circ}{\mathrm C}$.
After stripping from the mold, the ``semiproducts'' were cooled down in
an electric conveyer furnace in order to remove residual stress, thus
minimizing further deformations. Cooled semiproducts were then machined
using the classic operations used in the optical industry. After
quality testing, the reflecting surface was machined on vertical
milling machines. The ground mirror semiproducts were further submitted
to surface and geometrical checks. After washing and cleaning, the
surface was covered by the aluminium film ($95\:{\mathrm nm}$ thick) in
a vacuum tank. Mirrors were front-aluminized, which minimizes heat
losses by radiation during the night and therefore condensation. The
aluminium surface was then protected by a ${\mathrm Si O}_2$ layer
({$25\:{\mathrm nm}$ thick). The quality of the mirror surface was
checked in the following way: 80\% of the retro-reflected light from a
point source at the center of curvature falls within a circle of
$3\:{\mathrm mm}$ radius, which corresponds to an angular dispersion of
the direction of the normal  to the mirror at any point about its
theoretical value of $0.007^{\circ}$, half of the value required in the
design. 
\begin{figure}[htbp]
\epsfxsize=10cm
\centering
\leavevmode
\epsfbox{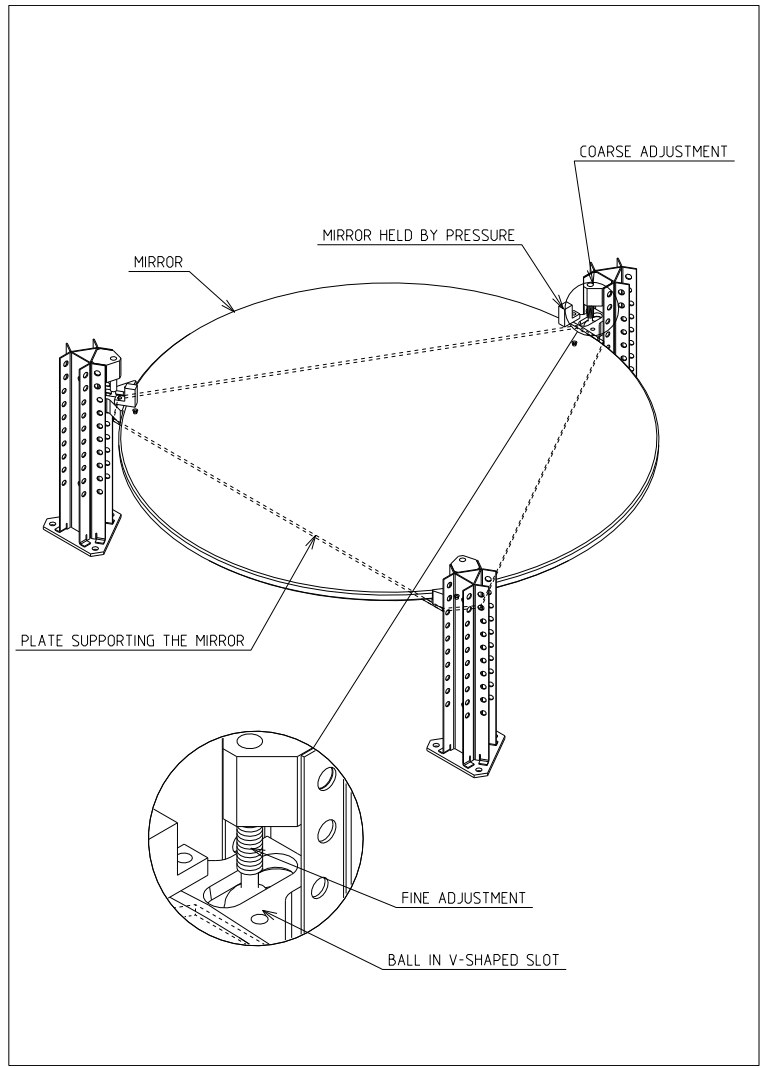}
\caption{Mirror-adjustment system}
\label{fig:mir}
\end{figure}

Each mirror is supported by a plate which is held by the mechanical
structure at three adjustable points; the adjustment systems are   
located in the free spaces between individual mirrors. Each such system
provides a coarse position adjustment and a fine orientation adjustment
(Fig.~\ref{fig:mir}).

Mirror orientations are tuned by means of three screws with spherical
heads, each of which is free to slide  along a V-shaped slot, thus
providing the degrees of freedom needed for dilatations and
adjustments. When all the mirror elements are correctly oriented, their
axes should converge to a common point $O$ located on the telescope
axis, namely the center of curvature of the central element $M_1$. The
adjustment of the mirror positions is checked in the following way
(Fig.~\ref{fig:reg}): mirrors are illuminated by a point-like  source
$S$ virtually located at point $O$; the light rays should be
retro-reflected by a lateral mirror element $M_2$ to the image point
$O'$, located on the axis of $M_2$, beyond the center of curvature $C$
of $M_2$. This beam is intercepted by an adjustable diaphragm $D$
located at $O$, and received by a video-camera focused on the mirrors
themselves; therefore, on the screen of the video-monitor,
well-oriented mirrors should appear illuminated with azimuthal
symmetry.  By progressively reducing the size of the diaphragm, finer
adjustments are obtained, eventually yielding an orientation accuracy
better than $0.008^{\circ}$. The optical tuning operation takes about
one day and was only needed twice during the first year of operation.
\begin{figure}[htbp]
\epsfxsize=10cm
\centering
\leavevmode
\epsfbox{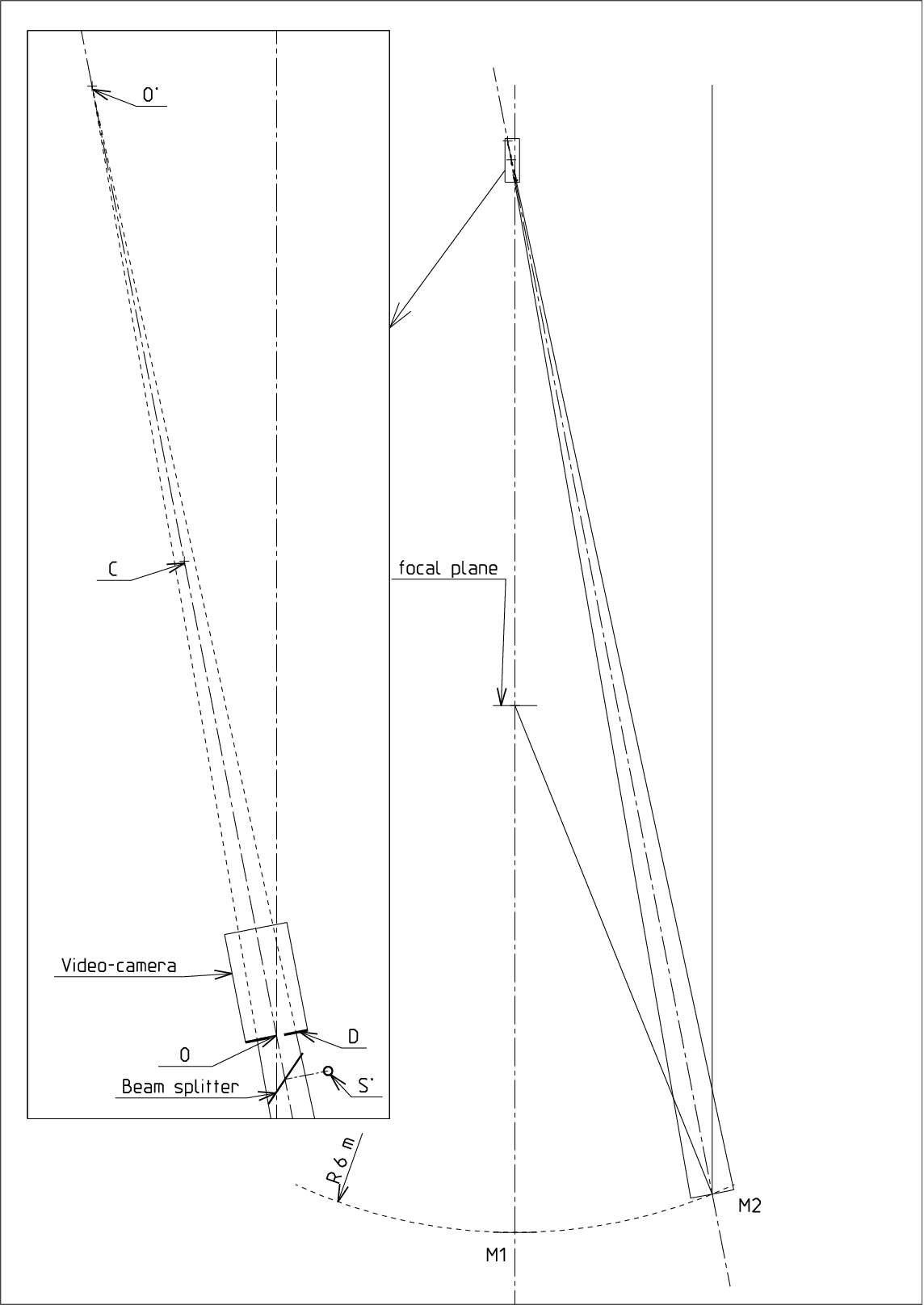}
\caption{Mirror-orientation tuning method. The box on the left
represents a zoom of the area around the center of curvature of $M_1$
indicated by a rectangle in the main figure.}
\label{fig:reg}
\end{figure} 
\begin{figure}[htbp]
\epsfxsize=10cm
\centering
\leavevmode
\epsfbox{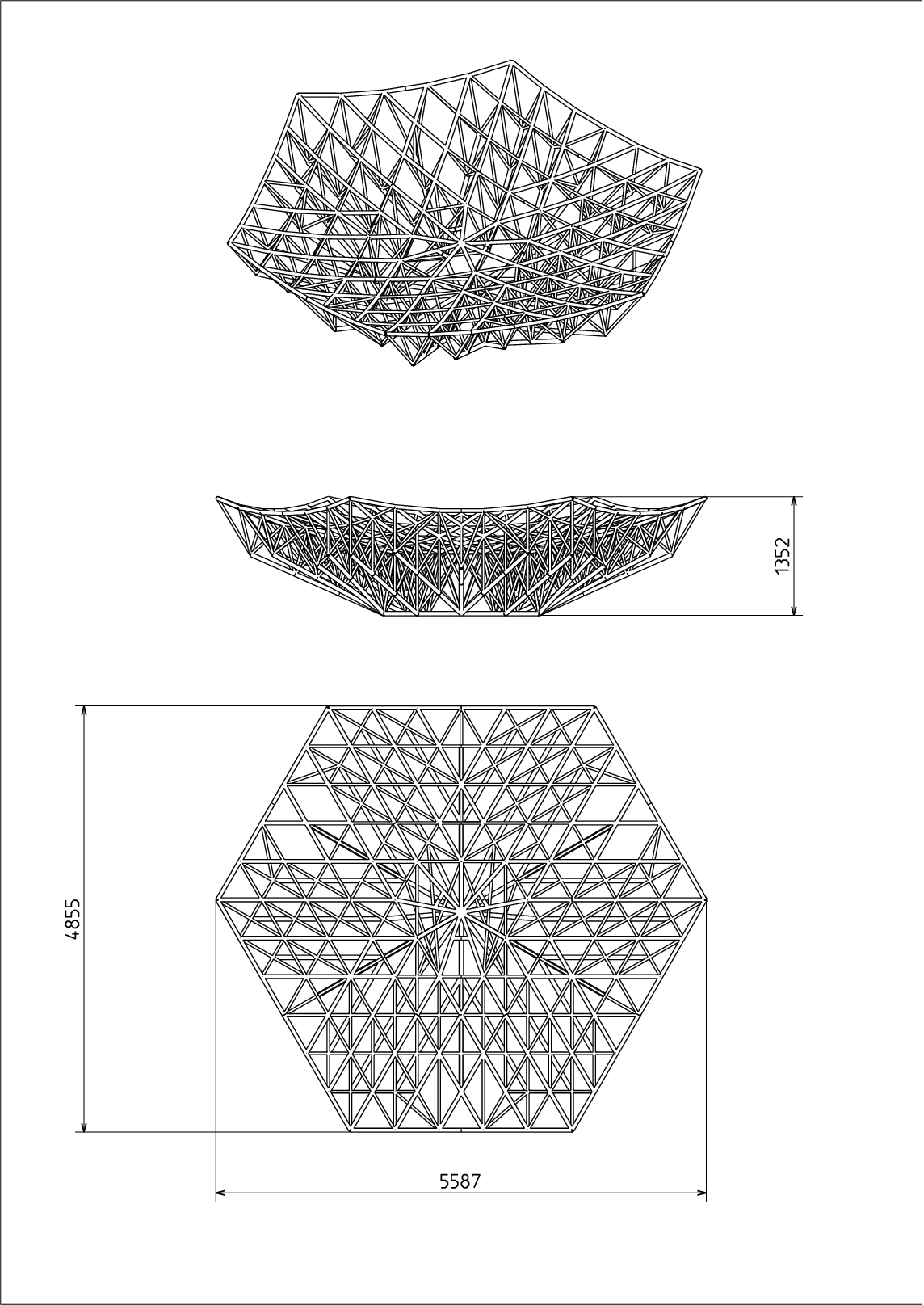}
\caption{The mirror support}
\label{fig:stru}
\end{figure}

\subsection { The mechanical structure }
The design of the mechanical structure was mainly constrained by the
following requirements: 
\begin{enumerate}
\item The whole structure should be light enough to be accommodated by
the  heliostat mount.
\item The mirror support should be rigid enough to keep mirror
orientations correct in all telescope positions.
\end{enumerate}
On the other hand, since the camera position is continuously monitored
as explained in section 3.3 below, small displacements of the focal
detector are tolerated. The whole structure was modelled using
finite-element analysis programs, thus optimizing the global shape as
well as discrete components to within the requirements. The study was
completed and finalized by an interactive modelling program ``{\small
EUCLID}'' \cite{Euclid} which allowed the manufacturing to be
contracted-out. The structure can be subdivided into three parts:
mirror support, camera support and connecting yoke, labeled  (1), (2),
and (3) respectively in Fig.~\ref{fig:vue}.
\begin{figure}[htbp]
\centering
\leavevmode
\epsfxsize=10cm
\epsfbox{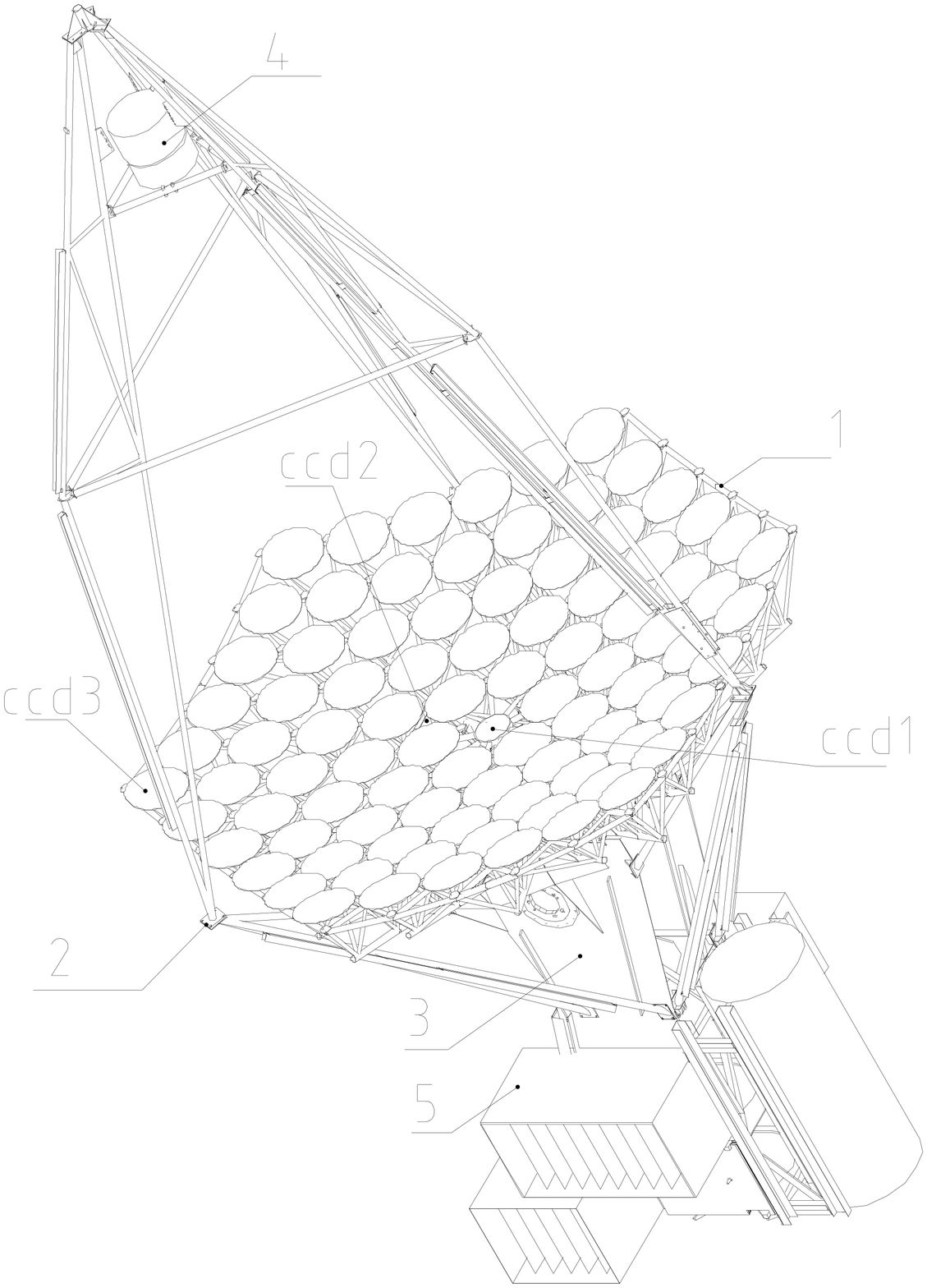}
\caption{General view of the mechanical structure; numbers are referred
to in the text}
\label{fig:vue}

\end{figure}

\subsubsection { The mirror support }
The mirror support consists of a reticulated structure of thin steel
tubes (Fig.~\ref{fig:stru}). The nodes on the front side are arranged
so as to match the adjustment systems  of the mirrors. The other nodes
and bars are laid out in order to get an isostatic structure in which
each bar works either in compression or in traction, thus providing
rigidity, lightness, and stability. This structure occupies a volume of
$5\times 4.8\times 1.3\:{\mathrm m}^{3}$ and weighs $700\:{\mathrm
kg}$. Particular care  was taken to weld the tubes together in such a
way that their axes  converge as closely as possible to the theoretical
positions of the nodes, an  essential condition for rigidity. An
annealing treatment was then performed in order to relieve stresses
induced by welding. The very small change in the dimensions observed
after annealing proved the high stability and rigidity of the mirror
support. The total load due to mirrors and adjustment systems amounts
to $600\:{\mathrm kg}$. In all telescope positions, this load induces
an angular deformation lower than $0.017^{\circ}$. These
characteristics were confirmed from the comparison of real images of
stars to images obtained by simulating the optics, including mirror
irregularities and residual misalignments. 

\subsubsection { The camera support }
The focal-plane detector (label 4 in Fig.~\ref{fig:vue}), located
$6\:{\mathrm m}$ away from the reflector, consists of the camera with
its 600 phototubes immediately followed by the trigger electronics. It
is held by a light support (label 2 in Fig.~\ref{fig:vue}), with
three arms, thus giving minimal optical shadowing while keeping a
sufficient safety coefficient to avoid buckling.  The load includes the
focal-plane detector itself ($110\:{\mathrm kg}$) and the cables
connecting the detector to the readout electronics ($250\:{\mathrm
kg}$). The camera support, coupled to the connecting yoke (label 3 in
Fig.~\ref{fig:vue}), is completely independent of the mirror support
(Fig.~\ref{fig:vuede}); the latter is therefore not submitted to
additional stresses. These different parts are indicated with the same
labels in Fig.~\ref{fig:vuede}. The small deformation of the tubes
leads to displacements of the focal detector perpendicular to the
telescope axis. In order to correct for this effect, the detector
position is continuously monitored by a {\small CCD} camera (label
{\small CCD1} in Fig.~\ref{fig:vue}) located on the mirror support
and yielding the positions of three photodiodes inserted in the
honeycomb structure holding the phototubes. The maximal shifting of the
focal-plane detector when going from the vertical to the horizontal
position of the telescope is $1.2\:{\mathrm cm}$, corresponding to
$0.12^{\circ}$ in the focal plane.
\begin{figure}[htbp]
\centering
\leavevmode
\epsfxsize=10cm
\epsfbox{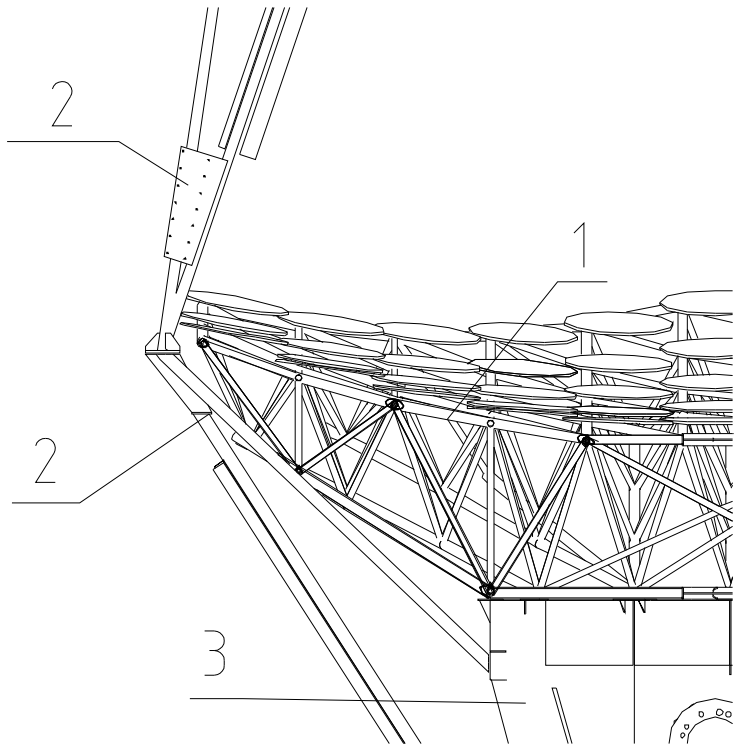}
\caption{Detailed view of the connection between parts labeled 1
(mirror support), 2 (camera support), and 3 (connecting yoke).}
\label{fig:vuede}
\end{figure}

\subsubsection { The connecting yoke }
The connecting yoke (label 3 in Fig.~\ref{fig:vue}), directly fixed
on the telescope  mount, provides the connection  between the mirror
support and the camera support. It also  holds the counterweight in
which all the readout electronics and power supplies (label 5 in
Fig.~\ref{fig:vue}) are accommodated. In this way, cables coming
from the focal-plane detector remain bound to the moving part of the
telescope and are not submitted to twisting. The connecting yoke allows
any elevation angle between $0^{\circ}$ and $90^{\circ}$ to be
reached. It covers the mount completely at zenith. The balance of the
system is adjusted so that the torque in elevation acts always in the
same direction, so as to have no play in the elevation gearing.

\subsection {Mechanical monitoring by CCD cameras}
\label{sec-ccd}
Three {\small CCD} cameras (referred to as {\small CCD1}, {\small
CCD2}, and {\small CCD3} respectively in Fig.~\ref{fig:vue}) fixed
on the mirror support, are used to monitor mechanical deformations and
to measure the pointing correction, i.e. the small shift between the
theoretical direction of the source aimed at and that of the telescope
axis. Camera {\small CCD1} monitors the position of the focal-plane
detector with respect to the reflector reference frame by means of
three photodiodes in the focal plane. Camera {\small CCD2}, located
close to the reflector center and co-aligned with its axis, and camera
{\small CCD3}, taking the place of an edge mirror element, both monitor
the star field surrounding the source.  Comparison of the images in
cameras {\small CCD2} and {\small CCD3} allows verification that the  
deformation of the mirror support remains within the limits expected
from the  finite-element modelling. The transformation which maps the
star field viewed by camera {\small CCD2} onto that viewed by the
focal-plane detector is obtained in dedicated periods of data-taking in
which a white screen is placed in the focal plane and the telescope is
continuously aimed at a bright star-field. The image formed on the
screen is viewed by camera {\small CCD1}, together with the three
photodiodes which monitor the position of the focal-plane detector.
These measurements allow the position of the optic axis in the focal
plane to be parameterized as a function of azimuth and elevation, using
a model which takes into account the gravitational bending of the
camera support and the slight misalignments of the azimuth and
elevation rotation axes.
\begin{figure}[htbp]
\centering
\leavevmode
\epsfxsize=10cm
\epsffile[40 160 530 660]{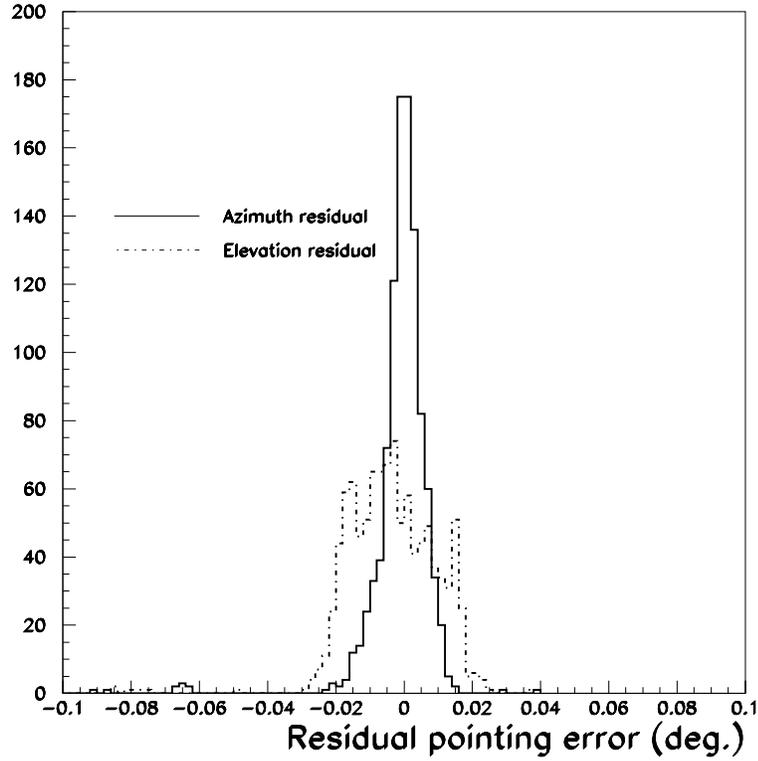}
\caption{Distribution of angular differences between the telescope
direction measured from {\small CCD} cameras and that predicted by the
model explained in the text.}
\label{fig:corr}
\end{figure}

Using this model together with data of cameras {\small CCD1} and
{\small CCD2} in normal data-taking periods, the source can be
localized in the field of view of the main camera with an accuracy of
$0.018^{\circ}$, i.e. better than $1/6$ of the pixel size, as shown in
Fig.~\ref{fig:corr}. This has been confirmed by using two independent
methods. In the first, the telescope was maintained at different fixed
orientations and the tracks of bright stars through the phototubes were
used to determine the true telescope position. In the second method,
the star field was directly identified from the readout of phototubes
during normal data taking; this is achievable with $0.12^{\circ}$
pixels provided that the field of view includes at least 4 stars with
magnitude lower than 7; in this case the direct mapping of the camera
reference frame onto the sky is obtained for short time-intervals
($\sim 1\:{\mathrm minute}$). All three methods give consistent results
in the determination of the source position in the field of view within
the  accuracy quoted above.

\section {The focal-plane detector}

The imaging camera, located in the focal plane 6 meters away from the
mirrors, provides a field of view of $3.1^{\circ}$ diameter covered by
small phototubes ($0.12^{\circ}$ diameter), extended to $4.8^{\circ}$
by two guard rings of larger tubes, as shown in  Fig.~\ref{fig:PM}.
All 600 phototubes are equipped with aluminized Winston cones.  
\begin{figure}[htbp]
\centering
\leavevmode
\epsfxsize=8cm
\epsffile[0 25 580 520]{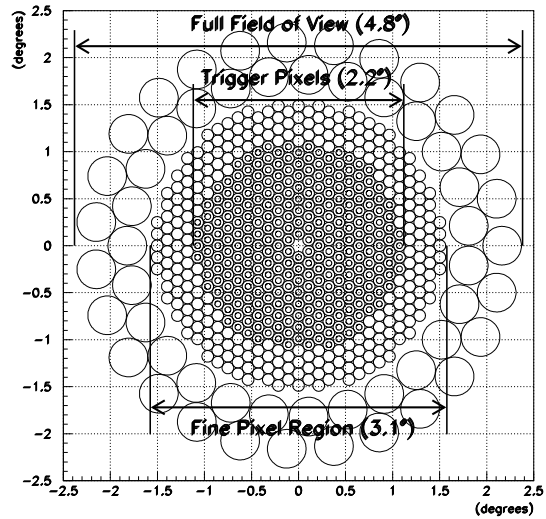}
\caption{Position of the 546 phototubes in the focal plane. The central
zone corresponds to the  288 phototubes involved in the trigger. The
field of view is extended to $4.8^{\circ}$ by two rings of 54 larger
phototubes.} 
\label{fig:PM}
\end{figure}

\subsection {Photomultipliers}

The inner phototubes are Hamamatsu R3878, which have the following
characteristics:
\begin{itemize}
\item a very small diameter of $11\:{\mathrm mm}$ to achieve the goal
of a high granularity for the matrix;
\item a very fast rise time ($<1\:{\mathrm ns}$) and narrow width
($\simeq1.4\:{\mathrm ns}$) to match the speed of the Cherenkov pulse;
\item good photoelectron resolution (see Fig.~\ref{fig:photo}) 
for an accurate photon count;
\item low dark-noise rate ($<150\:{\mathrm Hz}$ at a 0.3-photoelectron
threshold with a photo\-tube gain of $10^6$); 
\item {\small UV} glass to match the wavelength range of the Cherenkov
light taking account of atmospheric absorption.
\end{itemize}
 
A laboratory calibration was performed \cite{elec} which yielded the
gain as a function of high voltage, the single photoelectron peak's
mean value $Q$ and standard deviation $\sigma$ being obtained for each
phototube by a Gaussian fit.  On a large sample, the average value of
$\sigma / Q $ was found to be $0.43$ with an {\small RMS} of 0.04
(independent of the high voltage), with a peak/valley ratio of $\sim 2$ 
at a gain of $10^6$. Assuming a relation between the phototube gain
$G$ and the high voltage of the form $V=kG^{\alpha}$, the value of
$\alpha$ was measured for each phototube, allowing a determination of
the high voltage $V_6$ for the desired operating gain of $10^6$. The
average values were found to be $ \bar{\alpha} =  0.199$ and 
$\bar{V_6}=994\:{\mathrm V}$ and the corresponding {\small RMS}'s
$\Delta \alpha = 0.008$ and $\Delta V_6 = 56\:{\mathrm V}$. Because the
trigger zone of the camera requires a small transit time spread, the
phototubes were grouped by transit time for each sector, since a
$100\:{\mathrm V}$ variation entails a $600\:{\mathrm ps}$ shift. The 
conversion factors ({\small ADC} counts)/(photoelectron) are currently
measured every month using a pulsed {\small LED} at very low light
level; the single photoelectron spectrum is thus observed at a gain of
$3 \times 10^6$, allowing the peak to be well separated from the
pedestal (Fig.~\ref{fig:photo}) and providing a low enough fitting
threshold ($\simeq 0.3$ photoeletrons). In order to evaluate the
influence of the night-sky background on the trigger rate, phototubes
were illuminated through optical fibres with white light. The counting
rate variation as a function of the comparator threshold with
$G=10^{6}$ clearly shows two different zones with a break interpreted
as due to after-pulses (Fig.~\ref{fig:compt}).  Taking this into
account, it is found that an individual threshold around 
3~photoelectrons leads to a reduction of $10^3$ in the sky noise rate.
\begin{figure}[htbp]
\epsfxsize=10cm
\leavevmode
\centering
\epsffile[5 20 520 520]{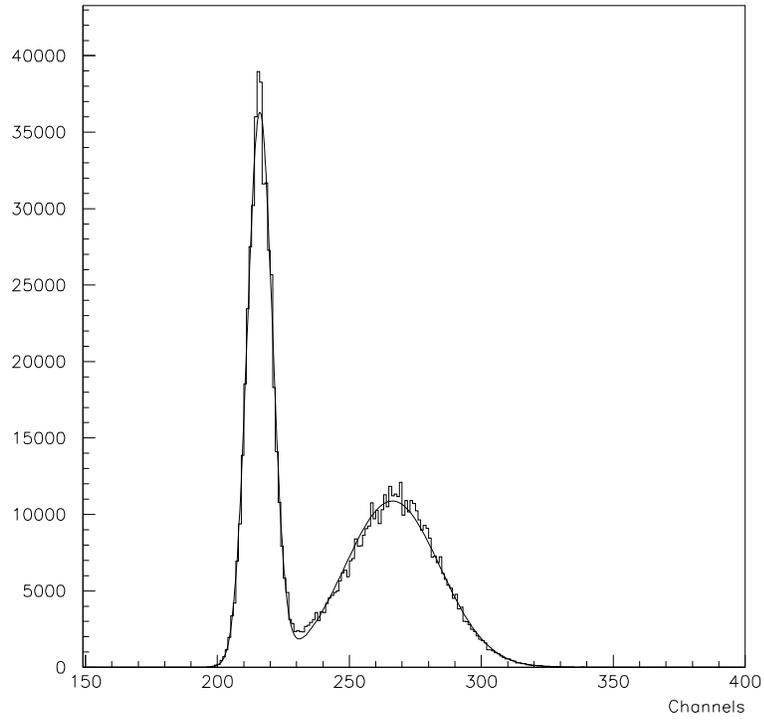}
\caption{Phototube calibration at an elevated high voltage:
Distribution of {\small ADC} counts.}
\label{fig:photo}
\end{figure}

The larger guard phototubes are Hamamatsu R6076 with a dia\-meter of  
$28.5\:{\mathrm mm}$, a fast rise time  ($<2.5\:{\mathrm ns}$) and a
photoelectron resolution comparable to that of the small phototubes.

\begin{figure}[htbp]
\epsfxsize=10cm
\leavevmode
\centering
\epsffile[5 20 370 370]{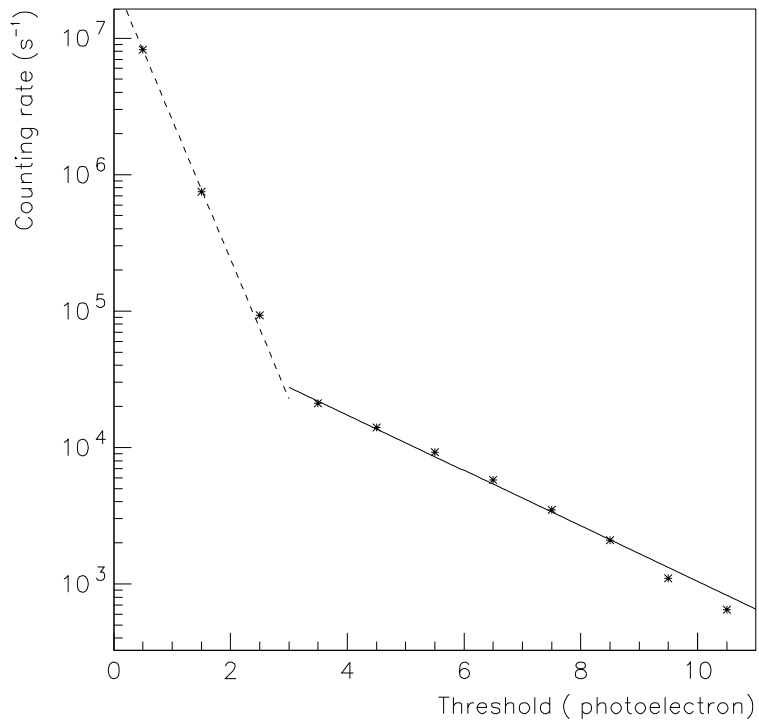}
\caption{Phototube counting rate as a function of comparator threshold
in units of photoelectrons.}
\label{fig:compt}
\end{figure}

\subsection {Winston collecting cones}
 
The Winston cones \cite{win} are an essential element of the camera
optics \cite{cones}. The dead-space between the photocathodes of the
phototubes is of the order of 65\% for the fine-pixel region. It can be
substantially reduced to approximately 10\% using reflective ``light
funnels'' to gather the photons falling between the phototubes. The
second major role of the cones is to block out much of the albedo light
coming from beyond the mirror, back-reflecting it away from the
photocathode. 
\begin{figure}[htbp]
\epsfxsize=8cm
\leavevmode
\centering
\epsffile[5 20 540 520]{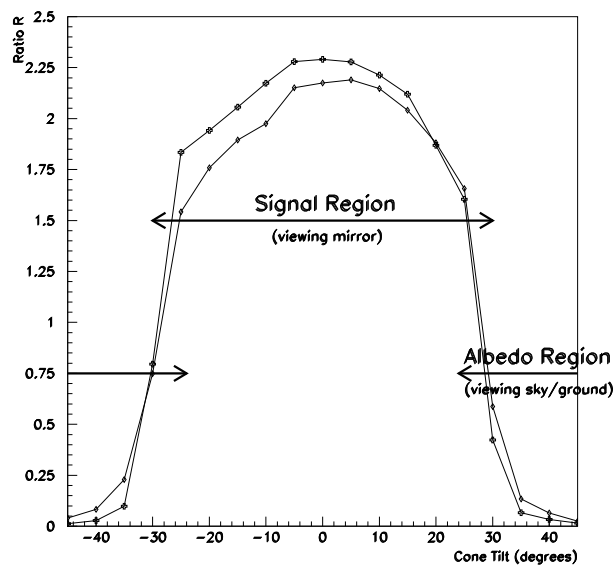}
\caption{Ratio $R$ of light-yields for a phototube with a Winston cone
to a bare phototube as a function of the tilt angle, as measured for
two Winston cones.}
\label{fig:fcones}
\end{figure} 

The Winston cones are milled from plastic, with three coats of varnish
applied to the interior to smooth the surface, on which are deposited a
reflective aluminium coat with a quartz overcoating. The entry diameter
of the cone is $12.8\:{\mathrm mm}$, and the exit diameter of the cone
is $6.25\:{\mathrm mm}$, with an interior form such that photons
arriving at angles greater than $30^\circ$ should be rejected while
those at lesser angles should arrive on the photocathode either
directly or after a single reflection. Since the exit of the diameter
of the cone is smaller than the photocathode  diameter ($\simeq
8\:{\mathrm mm}$), only the central part of the photocathode is used,
thus reducing the transit time dispersion and increasing the collection
efficiency. Measurements of the response of these Winston cones as a
function of the light incident angle are shown in
Fig.~\ref{fig:fcones}, from which it can be seen that the light from
the albedo region is greatly reduced, while about twice as many photons
coming from the mirror are collected  than with a bare phototube. The
reflectivity of the cone integrated with a diffuser reproducing the
shape of the mirror and seen under the same solid angle was measured at
different wavelengths. Averaging over the angle of incidence and over
the wavelength range of interest, the collection efficiency of the
cones was found to be $\sim 69\%$.

Equivalent Winston cones with an entry diameter of $42.0\:{\mathrm mm}$
and an exit diameter of $22.2\:{\mathrm mm}$ equip the guard
phototubes.
 
\section {Electronics}
\label{sec-elec}
\subsection {General architecture}

In order to minimize the use of long cables, the electronics was
designed in two parts:
\begin{itemize}
\item a very compact part, including the trigger logic, located just
behind the phototubes;
\item a second part located in the counterweight of the telescope
(label 5 in Fig.~\ref{fig:vue}), including the high voltage, the
readout electronics and the on-line computer.
\end{itemize}
Therefore, the whole electronics is mounted on the moving part of the
telescope, thus avoiding the torsion of cables. Digitized data are
transmitted to the control room by means of an optical fibre. The time
of each event is measured to an accuracy of $100\:{\mathrm ns}$ by a
clock based on the Global Positioning System ({\small GPS}). 
Fig.~\ref{fig:syn} shows the overall structure of the signal
processing.
\begin{figure}[htbp]
\epsfxsize=13cm
\leavevmode
\centering
\epsfbox{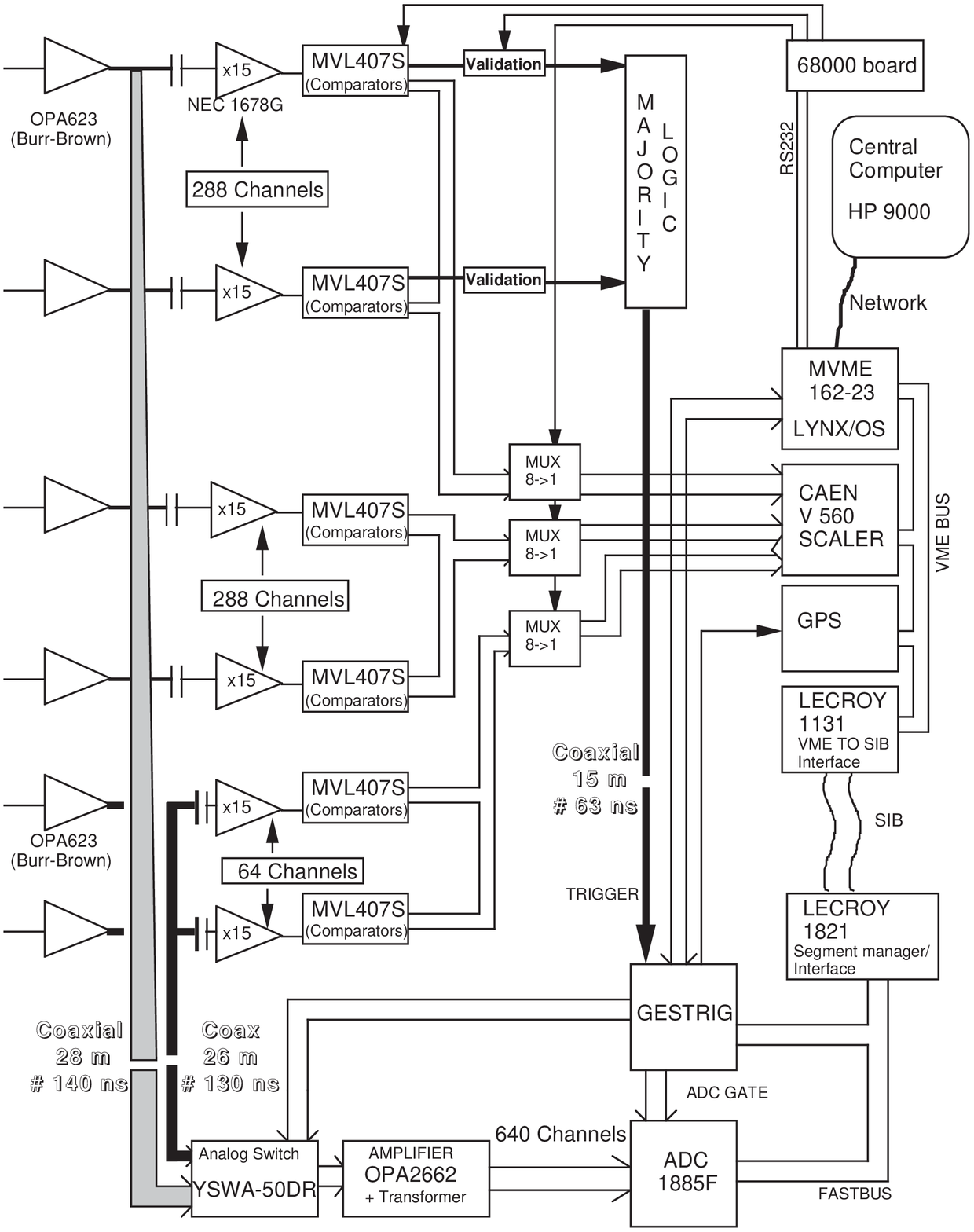}
\caption{General overview of the camera trigger, charge measurement and
data readout.}
\label{fig:syn}
\end{figure}

\subsection {The focal detector electronics}
\label{sec-trig}
The phototube pulse is first shaped by an {\small OPA}623 amplifier
which brings its width to $2.3\:{\mathrm ns}$ so as to take into
account fluctuations in the photon arrival times ($\simeq 2\:{\mathrm
ns}$) and to minimize the effect of the mirror asynchronism ($\simeq
1.6\:{\mathrm ns}$). After this front-end amplifier, the signal is
split into two electronics paths: the digital treatment to trigger the
camera and the analogue treatment (described in~\ref{sec-charge}) to
measure the charge at the anode of each phototube.

In order to have a good containment of shower images in the field of
view, the trigger zone was reduced to the inner 288 phototubes. The
trigger logic was designed to minimize random coincidences due to the
night-sky background, which has been measured to give $10^7$
photoelectrons per second in the phototubes. Given the known efficiency
of the detector, this is compatible with the generally quoted value of
$10^8\:{\mathrm photons.cm^{-2}sr^{-1}s^{-1}}$ \cite{Jelley}. A simple
majority coincidence logic involving the 288 phototubes would have led
to several tens of Hz of random triggers. The trigger zone was
therefore subdivided into 9 angular sectors of 48 phototubes, thus
reducing the combinatorial factor. This also minimizes the effect of
possible increases of the counting rates on several phototubes (e.g.,
if a star is momentarily located in a pixel) and reduces the rate of
triggering  hadronic showers. In order to avoid any loss of efficiency
at the sector boundaries, 16 phototubes from each sector overlap with
the adjacent sector (Fig.~\ref{fig:sect}).  The threshold level for
the phototubes and the number of tubes $k$ in coincidence in each
sector are adjustable under computer control. During the first year of
operation, the  phototube threshold has been fixed at the level of 3
photoelectrons, leading to a typical counting rate $\sim 10\:{\mathrm
kHz}$, and $k$ has been set to 4.  With such conditions, the random
coincidence rate is reduced to values well below $1\:{\mathrm Hz}$,
whereas cosmic-ray showers and muons contribute to about $15\:{\mathrm
Hz}$ at moderate zenith angles.
\begin{figure}[htbp]
\centering
\leavevmode
\epsfxsize=14.5cm
\epsfbox{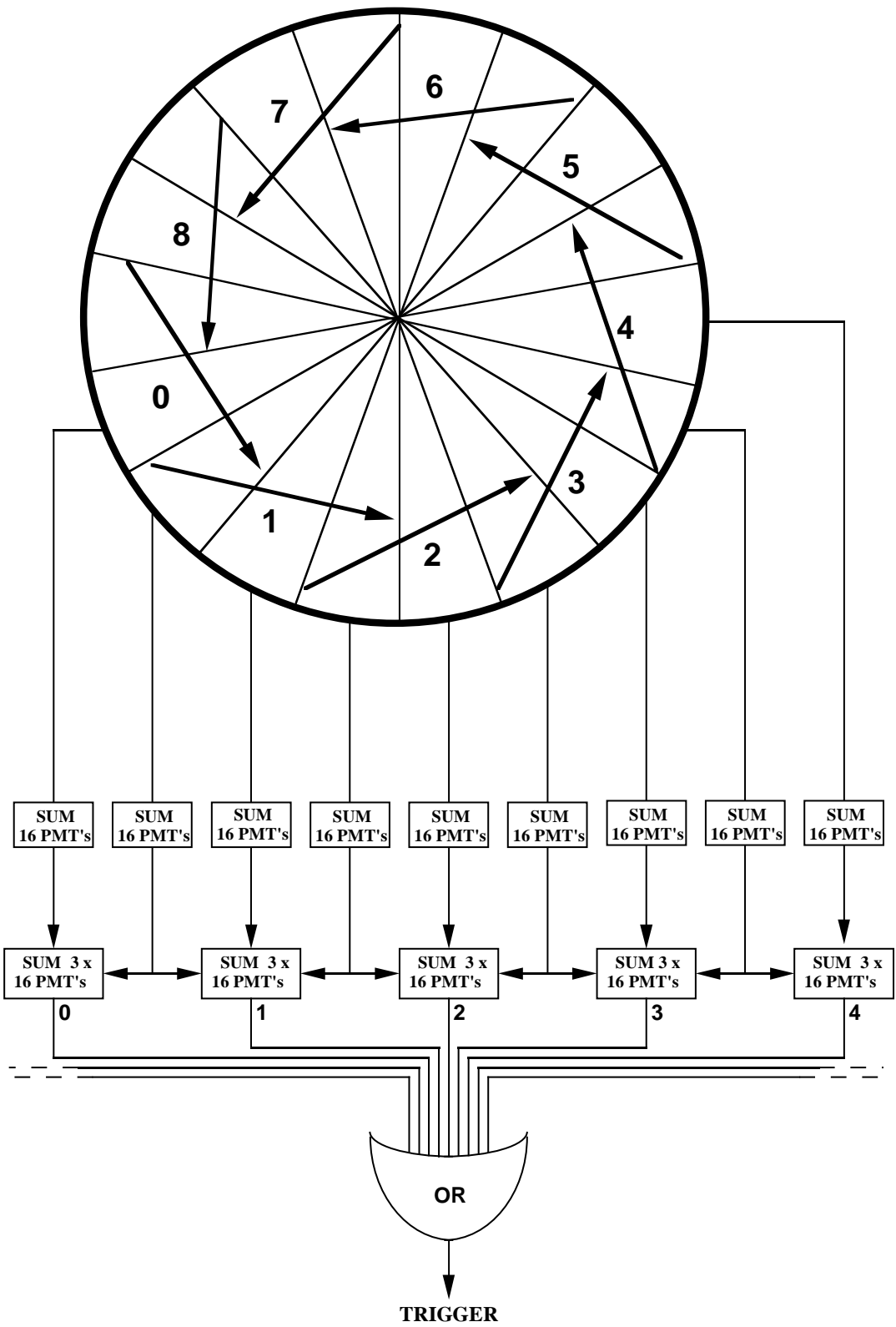}
\caption{Overview of the triggering logic with 9 ``overlapping'' sums.}
\label{fig:sect}
\end{figure}

To minimize the coincidence time-window as much as possible,
comparators instead of discriminators are used, thus taking the best
advantage of the signal rapidity. The signal is first amplified by 15
with a wide-band {\small NEC}1678 amplifier to allow a very low
threshold of the order of 1 photoelectron (at a gain of $10^6$). A sum
is performed over 16 comparators' outputs which is fed to the second
stage where an analogue sum over the 3 outputs corresponding to the
preceding partial sum is performed (Fig.~\ref{fig:sect}). This way, 9
sums over 3 sub-sectors are obtained. The majority logic condition (at
least $k$ pixels among 48 above the threshold) is carried out
independently for the 9 sums by a discriminator. The trigger is enabled
by an {\small OR} logic function over all sectors. 

Each comparator output is also sent to an 8-channel multiplexer
connected to $100\:{\mathrm MHz}$ scalers. The counts are integrated
for each phototube over 1 second, with a cycle duration of 8 seconds.
This allows the singles count rate to be recorded for each phototube in
order to:
\begin{itemize}
\item monitor the long-term and short-term variations of the night-sky
background;
\item detect the presence of a star in the field of view of a phototube
to automatically decrease its high voltage;
\item detect channels with anomalous counting rates due to electronic
noise, in order to remove them from the trigger logic and from the
physical analysis.
\end{itemize}
 
The compact electronics is cooled using 17 computer-controllable fans
inset in the camera box. The temperatures at 8 different locations
within the box are continuously monitored by the computer and recorded
in the data.
 
\subsection {Charge measurement}
\label{sec-charge}

The analogue signals as well as the trigger signal are carried to the
{\small ADC}'s located in the counterweight through long cables
(Fig.~\ref{fig:syn}). In order to minimize the night-sky background
contribution, the charge is integrated over a gate as narrow as
possible compatible with the width of the signal. Therefore, fast
analogue gates are used to replace the {\small ADC} gate of the LeCroy
1885F modules which has a minimum width of $65\:{\mathrm ns}$. With the
background rate due to the sky noise quoted above, only 0.1
photoelectrons are expected within the $12\:{\mathrm ns}$ gate. The
electronic noise is dominant over the night-sky background with an
{\small RMS} of 0.4 photoelectron in each pixel. Gates are opened
simultaneously  for all phototubes with each camera trigger. The
phototubes are so fast that practically all the charge is contained in
such a gate.

After being filtered by the gate, the analogue signal is sent to an
{\small OPA}2662 wide-band amplifier (Gain$\simeq$4). The differential
output enters a 15-bit LeCroy 1885 Fastbus {\small ADC} (96 channels)
with 2 different conversion scales of $50\:{\mathrm fC}$ per count for
pulses up to $200\:{\mathrm pC}$ and of $400\:{\mathrm fC}$ per count
above $200\:{\mathrm pC}$. In this way, the very wide dynamical range
necessary to cover a gamma-ray energy range from a few  hundreds of GeV
to a few tens of TeV is achieved. The conversion time for the {\small
ADC}'s is $350\:{\mathrm \mu s}$, but the dead-time induced by the full
readout chain for the 600 phototubes is $6.5\:{\mathrm ms}$.

A larger $44\:{\mathrm ns}$ gate is also generated by
computer-controlled triggers every 10 real events.  This larger
integration time allows a precise determination of the night-sky
background by fitting an adapted function to the {\small ADC} counts
histogram. The average rate of background photoelectrons during each 
30-minute running period is used to correct the pedestal determination
obtained on real data. This correction is necessary since pulses'
overshoots due to the effect of capacitive links induce a shift of
several {\small ADC} counts. 

\subsection{High Voltage control}

High voltages are controlled and monitored by means of {\small CAEN
SY}527 high-voltage boards  (24 channels, model A933K). Each phototube
can be addressed via a {\small CAENET} bus. Individual voltages can be
modulated in a range of $600\:{\mathrm V}$, which is sufficient to
lower the high voltage of a tube illuminated by a bright star or
showing a pathological behaviour.

\section {Data acquisition and monitoring software}

The data-acquisition system for the C{\small AT} imager must control a
number of subsystems, all programmed in C:
\begin{itemize}
\item Imaging camera, including phototube high voltage control, {\small
ADC} and {\small GPS} readout, scalers readout, trigger logic
(section~\ref{sec-elec}), implemented on a Motorola 68020 processor.
\item Telescope tracking control (section~\ref{sec-setup}), implemented
on a {\small PC}.
\item {\small CCD} monitoring of telescope position
(section~\ref{sec-ccd}), implemented on a {\small PC}.
\end{itemize}
The three subsystems are connected over the network to a central
machine ({\small HP} 9000). In order to control this ensemble, the
``General Application Server'' ({\small GAS}) was developed. This
model, described in \cite{Gas}, is a flexible architecture for
experiment control, which is implemented using standard {\small TCP/IP}
services and {\small POSIX}-compliant code. In this way, portability,
maintainability, and re-usability goals are met.

In this model, each subsystem is controlled by a Control client and
monitored by one or more Data clients (Fig.~\ref{fig:acq}). These
clients are connected together via a {\small GAS} Server which launches
and communicates with the subsystem's Remote Control Task. The {\small
GAS} servers, Control clients, Remote tasks, and Data clients can run
anywhere on the net.   As each is an independent process communicating
with the others by {\small TCP/IP}, they are more easily modifiable and
debugging is simplified.
 
\vskip 5mm
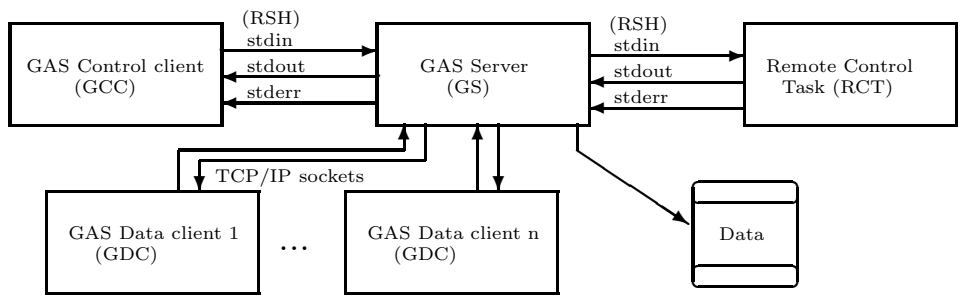
\begin{figure}[htbp]
\setlength{\unitlength}{0.00061200in}%
\begingroup\makeatletter\ifx\SetFigFont\undefined
\def\x#1#2#3#4#5#6#7\relax{\def\x{#1#2#3#4#5#6}}%
\expandafter\x\fmtname xxxxxx\relax \def\y{splain}%
\ifx\x\y   
\gdef\SetFigFont#1#2#3{%
  \ifnum #1<17\tiny\else \ifnum #1<20\small\else
  \ifnum #1<24\normalsize\else \ifnum #1<29\large\else
  \ifnum #1<34\Large\else \ifnum #1<41\LARGE\else
     \huge\fi\fi\fi\fi\fi\fi
  \csname #3\endcsname}%
\else
\gdef\SetFigFont#1#2#3{\begingroup
  \count@#1\relax \ifnum 25<\count@\count@25\fi
  \def\x{\endgroup\@setsize\SetFigFont{#2pt}}%
  \expandafter\x
    \csname \romannumeral\the\count@ pt\expandafter\endcsname
    \csname @\romannumeral\the\count@ pt\endcsname
  \csname #3\endcsname}%
\fi
\fi\endgroup
\begin{picture}(8123,2835)(2150,-4408)
\thicklines
\put(8913,-2178){\vector(-1, 0){1350}}
\put(8913,-2403){\vector(-1, 0){1350}}
\put(7788,-1908){\makebox(0,0)[lb]
{\smash{\SetFigFont{7}{8.4}{rm}stdin}}}
\put(7788,-2133){\makebox(0,0)[lb]
{\smash{\SetFigFont{7}{8.4}{rm}stdout}}}
\put(7788,-2358){\makebox(0,0)[lb]
{\smash{\SetFigFont{7}{8.4}{rm}stderr}}}
\put(8913,-3123){\oval(900,180)}
\put(8912,-3842){\oval(900,180)}
\put(4412,-1907){\makebox(9.0714,13.6071){\SetFigFont{10}{12}{rm}.}}
\put(5762,-2132){\makebox(9.0714,13.6071){\SetFigFont{10}{12}{rm}.}}
\put(7472,-2537){\line( 0,-1){225}}
\put(7472,-2762){\vector( 3,-2){955.385}}
\put(8463,-3123){\line( 0,-1){675}}
\put(9363,-3123){\line( 0,-1){675}}
\put(4412,-1907){\vector( 1, 0){1350}}
\put(5762,-2132){\vector(-1, 0){1350}}
\put(5762,-2357){\vector(-1, 0){1350}}
\put(6616,-3121){\vector( 0, 1){585}}
\put(6796,-2536){\vector( 0,-1){585}}
\put(4051,-3121){\line( 0, 1){360}}
\put(4051,-2761){\line( 1, 0){1935}}
\put(5986,-2761){\vector( 0, 1){225}}
\put(6166,-2536){\line( 0,-1){315}}
\put(6166,-2851){\line(-1, 0){1935}}
\put(4231,-2851){\vector( 0,-1){270}}
\put(2614,-2539){\framebox(1800,855){}}
\put(5763,-2538){\framebox(1800,855){}}
\put(7563,-1953){\vector( 1, 0){1350}}
\put(8913,-2538){\framebox(1800,855){}}
\put(5671,-3526){\makebox(0,0)[lb]
{\smash{\SetFigFont{7}{8.4}{rm}GAS Data client n}}}
\put(2929,-3979){\framebox(1800,855){}}
\put(5494,-3979){\framebox(1800,855){}}
\put(4637,-1862){\makebox(0,0)[lb]
{\smash{\SetFigFont{7}{8.4}{rm}stdin}}}
\put(4637,-2087){\makebox(0,0)[lb]
{\smash{\SetFigFont{7}{8.4}{rm}stdout}}}
\put(4637,-2312){\makebox(0,0)[lb]
{\smash{\SetFigFont{7}{8.4}{rm}stderr}}}
\put(8688,-3528){\makebox(0,0)[lb]
{\smash{\SetFigFont{7}{8.4}{rm}Data}}}
\put(6121,-2086){\makebox(0,0)[lb]
{\smash{\SetFigFont{7}{8.4}{rm}GAS Server}}}
\put(6346,-2266){\makebox(0,0)[lb]
{\smash{\SetFigFont{7}{8.4}{rm}(GS)}}}
\put(7741,-1726){\makebox(0,0)[lb]
{\smash{\SetFigFont{7}{8.4}{rm}(RSH)}}}
\put(4366,-3031){\makebox(0,0)[lb]
{\smash{\SetFigFont{7}{8.4}{rm}TCP/IP sockets}}}
\put(4591,-1681){\makebox(0,0)[lb]
{\smash{\SetFigFont{7}{8.4}{rm}(RSH)}}}
\put(2746,-2086){\makebox(0,0)[lb]
{\smash{\SetFigFont{7}{8.4}{rm}GAS Control client}}}
\put(3151,-2266){\makebox(0,0)[lb]
{\smash{\SetFigFont{7}{8.4}{rm}(GCC)}}}
\put(9091,-2086){\makebox(0,0)[lb]
{\smash{\SetFigFont{7}{8.4}{rm}Remote Control}}}
\put(9226,-2266){\makebox(0,0)[lb]
{\smash{\SetFigFont{7}{8.4}{rm}Task (RCT)}}}
\put(4908,-3618){\makebox(0,0)[lb]
{\smash{\SetFigFont{14}{16.8}{rm}...}}}
\put(3331,-3706){\makebox(0,0)[lb]
{\smash{\SetFigFont{7}{8.4}{rm}(GDC)}}}
\put(5896,-3706){\makebox(0,0)[lb]
{\smash{\SetFigFont{7}{8.4}{rm}(GDC)}}}
\put(3106,-3526){\makebox(0,0)[lb]
{\smash{\SetFigFont{7}{8.4}{rm}GAS Data client 1}}}
\end{picture}
\caption{The {\small GAS} Model used by each of the data-acquisition
subsystems}
\label{fig:acq}
\end{figure} 
\vskip 5mm
 
The acquisition is orchestrated by a ``Master Run Controller'', which
loads a separate {\small GAS} Control client for each subsystem (the
Control client being specific to that subsystem), and synchronizes
commands to open and close data files for the subsystems.   As the
Control clients run as sub-programs of the Master, common information
(e.g., source position, high-voltage status) is communicated via global
variables.

Each {\small GAS} Control client launches an instance of the {\small
GAS} server (by the Remote Shell {\small TCP/IP} service), and commands
it to launch the Remote Control Task for its subsystem on the
appropriate machine (also by remote shell).   The use of the standard
POSIX remote shell allows communication with the process launched by
{\small POSIX}'s standard {\em stdin}, {\em stdout}, and {\em stderr}
{\small I/O} channels. Typically, the Remote Control task is run on a
machine near the subsystem hardware.   The configuration of the
subsystem is stored in the {\small GAS} Server by the Control client
and passed to any relevant Data client which connects to it.  The
Control client sends commands to the Remote Control Task via the
{\small GAS} Server, which simply passes on the commands and returns
acknowledgements and alarms from the Remote task to the Control client. 
The data from the Remote task are sent to the {\small GAS} Server,
which stores them on disk if  requested by the Control client.  The
data files from each subsystem are combined off-line and translated
into F-Pack format \cite{fpack}.

Specific Data clients are used for monitoring of data quality, detector
state, et cetera.  For example, for the Imaging camera subsystem, Data
clients monitor the actual high voltage values,  scalers rates,
temperatures in the camera box, and provide also a sample of the
Cherenkov images. These Data clients can be launched either
automatically by the Master Run Controller at  start-up or manually by
the operator at any time during data-acquisition.   A data client
searches through the running {\small GAS} servers to find the one which
is connected to the desired subsystem. When found, the Data client
receives a copy of the configuration buffer from the {\small GAS}
Server and can then request a specific type of  data sample (high
voltages, scalers, etc.).  The Data clients communicate with the Server
using standard {\small TCP/IP} sockets.
  
The most complex subsystem to control is the Imaging camera. The Remote
control tasks for the Imaging camera run a Motorola 68020 processor
running under Lynx-{\small OS} mounted in the counterweight of the
telescope, close to the electronics. These Remote Control tasks are
implemented using grammar-interpreters which execute the control
statements coming from {\em stdin} and which output data to {\em
stdout} and acknowledgements to {\em stderr}. Thus, the control
statements are expressed in a readable and easy-to-understand language.
This approach facilitates their stand-alone execution and debugging:
the Control task can be launched locally, redirecting outputs to files
or filters, and the control statements typed to the input to analyze
hardware response and/or data outputs.
 
The readout tasks are {\small POSIX}-compliant threads. To enable
readout, the Remote Control task launches these threads, which output
data when triggered. The access to the output channel is a shared
resource, so a mutual exclusion is implemented using standard mutex
functions.
 
For the C{\small AT} Imaging telescope, the Master Run Controller,
Control clients,  and Data clients have been implemented in Lab{\small
VIEW}, a graphical programming language which allows user interface to
be easily developed. The {\small GAS} server itself is a small C
program ($\simeq$1000 lines) and a C-shell script. In this script the
Remote Control task is launched using the standard remote shell
command, and its outputs are piped to the {\small GAS} Server input.  
The use of this highly-modular architecture allowed the
data-acquisition to be  developed rapidly and should permit the other
experiments on the site to be integrated.
 
\section {Conclusion}

With the trigger conditions explained in section \ref{sec-trig}, random
coincidences due to the night-sky background have a negligible
contribution and the trigger rate, essentially due to hadronic showers
and single muons, is $\sim 15\:{\mathrm Hz}$ at moderate zenith angles
($\leq 30^{\circ}$).  The rate due to single muons---obtained when the
sky is overcast---amounts to $6\:{\mathrm Hz}$. At this threshold, the
rather moderate background-event rate is due to a primary rejection by
the trigger hardware.

From the observations of the Crab nebula between November 1996 and
March 1997, a preliminary estimate of the energy threshold ($\sim
250\:{\mathrm GeV}$)  has been given \cite{durb}. It may be possible to
lower the threshold further as the data acquisition system should be
able to cope with increased trigger rates. During the first year of
operation, three sources have been detected: the Crab nebula
\cite{durb},  Markarian 501 \cite{d501} and Markarian 421.

\begin{ack}
The authors wish to thank the French national institutions IN2P3/CNRS
and DAPNIA/DSM/CEA for supporting and funding the C{\small AT} project.
The C{\small AT} telescope was also partly funded by the
Languedoc-Roussillon region and the Ecole Polytechnique. The authors
also wish to thank Electricit\'e de France for making available to them
equipment at the former solar plant ``Th\'emis'' and allowing the 
building of the new telescope and its hangar. They are grateful to the
French  and Czech ministries of Foreign Affairs for providing grants
for physicists' travel and accommodation expenses.
\end{ack}

\end{document}